\pdfoutput=1
\documentclass[sigconf,nonacm]{acmart}

\renewcommand\footnotetextcopyrightpermission[1]{}

\usepackage{booktabs}
\usepackage{enumitem}
\usepackage{listings}
\usepackage{xcolor}
\lstdefinestyle{shiftypython}{
  language=Python,
  basicstyle=\footnotesize\ttfamily,
  keywordstyle=\color[rgb]{0.00,0.44,0.13}\bfseries,
  stringstyle=\color[rgb]{0.73,0.13,0.13},
  commentstyle=\itshape\color[rgb]{0.25,0.50,0.50},
  showstringspaces=false,
  breaklines=true,
  columns=fullflexible,
  keepspaces=true,
  aboveskip=0pt,
  belowskip=0pt,
  xleftmargin=0pt,
  frame=none,
}
\usepackage{tabularx}
\usepackage{tikz}
\usepackage{forest}
\usepackage{xspace}
\usetikzlibrary{arrows.meta,positioning}

\newcommand{\shifty}{\textsc{Shifty}\xspace}
\newcommand{\systemname}{\shifty}

\title{Evidence-Carrying Validation for Knowledge Graphs}

\author{Gabe Fierro}
\affiliation{%
  \institution{Colorado School of Mines}
  \city{Golden}
  \state{Colorado}
  \country{USA}
}
\email{gtfierro@mines.edu}

\begin{document}

\begin{abstract}
Programs that consume a knowledge graph they do not maintain, such as applications, authoring platforms, and LLM agents, need to know whether the graph contains the information their task requires.
Validating the graph against a schema can answer this question, but existing validation interfaces usually return a conformance bit or failure-oriented report without identifying why checks pass or the partial matches behind failures.
We present an \emph{evidence-carrying validation} interface: every selected node--shape check returns either a satisfaction trace or failure witness.
These are mutually recursive objects that retain constraints, cardinality decisions, paths, and supporting triples.
We implement this interface in \systemname, an experimental SHACL validator.
Against two real-world shape graph corpora, materializing all-pair evidence costs a median $1.54$--$2.07\times$ conformance-only validation.
A case study then shows how programs combine passing and failing evidence to diagnose missing information and guide repair.
\end{abstract}

\maketitle

\section{Introduction}
\label{sec:introduction}

Knowledge graphs increasingly serve programs that do not maintain them: applications configure themselves from graph contents, platforms author and extend graphs, and LLM agents query and repair them.
The constraints on a graph evolve alongside it: maintainers add and revise requirements in the graph's ontology, and new use cases demand structure and metadata the graph does not yet carry.
Shapes make these expectations testable: validation of a graph against a shape determines \emph{semantic sufficiency}---whether the current graph contains enough structure and metadata to configure and execute an application~\cite{fierro_application_2022}.
Semantic models already play this role in building-control and water-treatment systems, where applications discover sensors and contextualized operational data across heterogeneous sources~\cite{balaji_brick_2018,roth_supervisory_2022,saka_acquirium_2025}.

Acting on a validation result requires specifics.
On success, the application needs the concrete nodes and graph paths that satisfy the contract.
On failure, an operator or repair program needs the reachable candidates and missing obligations that explain what the graph lacks.
Existing validation interfaces provide neither view completely.
SHACL localizes violations but does not enumerate passing focus nodes or require nested evidence sufficient to justify each failure~\cite{noauthor_shapes_2017}.
An absent SHACL result is consequently ambiguous to a consumer: the node may have passed, its target selector may have missed it, or the evaluator may never have checked it.
Even an explicit pass does not identify the validating subgraph that an application should consume.

We propose \emph{evidence-carrying validation}: a validator returns a \emph{satisfaction trace} when a node satisfies a shape and a \emph{failure witness} when it does not.
Both objects follow a common shape algebra and become mutually recursive through negation: a trace contains its child's witness or a witness contains its child's trace.
Each object retains the selected focus, evaluated constraint, decisive Boolean and cardinality evidence, and the concrete triples supporting graph paths.
The result is a pruned, machine-interpretable justification.
We make the following contributions:
\begin{itemize}[noitemsep]
  \item a validation result interface that exposes target coverage, both result polarities, result-relative complete justification, and concrete graph support;
  \item a polarity-aware evidence semantics with mutually recursive failure witnesses and satisfaction traces; and
  \item an implementation that exposes typed Python result objects, preserves authored identities through normalization, and materializes evidence completely or on demand.
\end{itemize}
\vspace{-1em}

\section{The Missing Validation Interface}
\label{sec:reports}

Most graph validation interfaces discard the information used to compute conformance.
A validation report may identify a violation while omitting which other nodes were selected, what made their checks pass, or how closely a candidate matched before it failed.
Recovering these facts can be expensive, requiring rerunning or even reverse-engineering the validator.
Instead, validation should preserve what the validator learned, and present that information in a form another program can use.

\begin{table*}[ht]
  \caption{Coverage and evidence exposed by validation interfaces.
  The PG-Schema language defines conformance but no result vocabulary, so its author-maintained BPG validator appears separately.
  Complete justification recursively includes every child needed to establish a particular result.}
  \label{tab:interfaces}
  \vspace{-1em}
  \scriptsize
  \setlength{\tabcolsep}{3pt}
  \begin{tabularx}{\textwidth}{@{}
      >{\raggedright\arraybackslash}p{0.14\textwidth}
      >{\raggedright\arraybackslash}p{0.10\textwidth}
      >{\raggedright\arraybackslash}p{0.09\textwidth}
      >{\raggedright\arraybackslash}p{0.10\textwidth}
      >{\raggedright\arraybackslash}p{0.13\textwidth}
      >{\raggedright\arraybackslash}p{0.14\textwidth}
      >{\raggedright\arraybackslash}X@{}}
    \toprule
    & Selected checks & \multicolumn{2}{c}{Observable results} & \multicolumn{2}{c}{Justification} & Data grounding \\
    \cmidrule(lr){3-4}\cmidrule(lr){5-6}
    Interface & Coverage & Passing & Failing & Passing & Failing & \\
    \midrule
    SHACL report~\cite{noauthor_shapes_2017}
      & Not enumerated & Report-wide status & Violation records & None & Optional nested detail & RDF terms \\
    ShEx ShapeMap~\cite{prudhommeaux_shex_2019,shapemap_2019}
      & All requested pairs & Per-pair status & Per-pair status & None standardized & Text or app-defined & RDF terms \\
    JSON Schema verbose output~\cite{jsonschema_core_2020}
      & Evaluated locations & Passing records & Failing records & Complete hierarchy & Complete hierarchy & Document locations \\
    PG-Schema specification~\cite{angles_pg_schema_2023}
      & \multicolumn{6}{l}{Conformance semantics; no standardized validation result vocabulary} \\
    BPG PG-Schema validator~\cite{tomaszuk_bpg_pgs_validator_2026}
      & Element counts & Report-wide status & Issue records & None & Local issue only & Elements/properties \\
    Evidence-carrying validation
      & All selected nodes & Per-focus result & Per-focus result & Complete trace & Complete witness & RDF triples \\
    \bottomrule
  \end{tabularx}
  \vspace{-1em}
\end{table*}

A validation result object forms the boundary between the validator and its consumers.
Its contents determine whether a consumer can explain a result, extract conforming data, diagnose target selection, or propose a repair without invoking the validator again.
When a report contains no failure for a node, a consumer must be able to distinguish a passing check from a missed target or an unevaluated check.
The report therefore needs to enumerate the selected node--shape checks and attach an explicit passing or failing record to each one.
We call this enumeration the report's \emph{coverage}.

The result record should also preserve the constraint evaluations that establish the pass or failure and identify the input triples that those evaluations used.
We call the first part the result's \emph{justification} and the second its \emph{data grounding}.

A justification is \emph{complete} relative to one passing or failing result.
It includes every child required to establish that result:
e.g., every child of a successful conjunction, or every branch of a failed disjunction.
Justifications are ideally symbolic; prose can summarize a result for human operators, but programs need access to typed constraints, values, and paths without parsing explanatory text.

Table~\ref{tab:interfaces} shows that existing interfaces omit these properties independently.
SHACL gives a consumer useful coordinates for a violation (focus node, result path, source shape, and constraint component) and may attach nested detail through \texttt{sh:detail}.
However, it does not require that detail to justify the failure completely, and it returns no corresponding record for a passing focus node.
The absence of a result therefore remains ambiguous.

ShEx ShapeMaps close that gap by recording each requested pair's polarity, proving that the validator selected and checked the node.
They do not standardize graph support for passing nodes or structured failure reasons for violating nodes.
Outside graph validation, JSON Schema's verbose output provides a useful precedent: it records both passing and failing evaluations in a complete hierarchy grounded in document locations~\cite{jsonschema_core_2020}.
The PG-Schema specification defines graph conformance without a result vocabulary~\cite{angles_pg_schema_2023}; its author-maintained BPG validator reports checked-element counts and failure-oriented issues with only local information~\cite{tomaszuk_bpg_pgs_validator_2026}.

Our proposed evidence-carrying validation interface preserves the information that these interfaces expose separately.
Every selected node receives a satisfaction trace or a failure witness.
The result records constraint structure, focus and value nodes, alternatives, cardinality counts, traversed paths, and matched triples.
A consumer can traverse the evidence and project matched values, missing obligations, offending values, and supporting triples for downstream use.
The next section gives these justifications a logical structure that preserves only the evidence required for a result.
Algebraic simplification can make a justification smaller still, but may break its correspondence to named source constraints.

\section{Evidence Semantics for Graph Validation}
\label{sec:witnesses}

A validation result remains useful outside the validator only if another program can interpret its justification without reproducing the validator's control flow.
The validation evidence therefore needs a stable logical structure for both passing and failing checks, together with the input triples that connect each focus node to the values used in the check.

\subsection{A Shared Shape Algebra}

Validation-language syntax cannot provide this stable structure.
Equivalent constraints may use different components, and an engine may reorganize them during evaluation.
We therefore assign each trace and witness a constraint from the common shape algebra formalized by Ahmetaj et al.~\cite{ahmetaj_common_2025}.
Shapes and paths are:
\[
\begin{split}
  \pi &::= \mathit{id}\mid p\mid\pi^{-}\mid\pi\mathbin{/}\pi
          \mid\pi\mathbin{\cup}\pi\mid\pi^{*},\\
  \varphi &::= \top\mid\mathit{test}\mid\mathit{opaque}(\omega)\mid\neg\varphi
          \mid\varphi\land\varphi\mid\varphi\lor\varphi\\
       &\phantom{::=}\mid \exists_{\geq n}\pi.\varphi
          \mid \exists_{\leq n}\pi.\varphi .
\end{split}
\]
Here $n\in\mathbb{N}$ is a cardinality bound and $\top$ is the shape that always passes.
A predicate $p$ denotes one RDF edge, $\mathit{id}$ is the zero-length identity path, $\pi^{-}$ reverses a path, $/$ composes paths, $\cup$ selects alternatives, and $*$ repeats a path zero or more times.
$\mathit{test}$ ranges over atomic constraints, such as \texttt{sh:hasValue}, for which the validator can report observed and expected values.
$\mathit{opaque}(\omega)$ represents an extension constraint $\omega$ for which the validator can report only pass or fail.
Other language-specific constraints can appear as atomic tests when the validator can expose their inputs.

\subsection{Satisfaction Traces and Failure Witnesses}

\begin{sloppypar}
The validator applies each named shape to the nodes selected by its target expression.
For every selected node $v$, the validation result contains either a satisfaction trace $S$ or a failure witness $F$:
\[
  E_v ::= \mathsf{Satisfaction}(S)\mid\mathsf{Failure}(F).
\]
If a shape selects no nodes, the result records an empty selection rather than a failure.
For graph $G$, node $v$, and shape $\varphi$, let 
  $\mathsf{Sat}(G,v,\varphi)$
  and
  $\mathsf{Fail}(G,v,\varphi)$
denote the sets of valid satisfaction traces and failure witnesses for that check.
If $G,v\models\varphi$, then $\mathsf{Sat}(G,v,\varphi)$ is nonempty and $\mathsf{Fail}(G,v,\varphi)$ is empty; the reverse holds for $G,v\not\models\varphi$.
A validation judgment may have several valid explanations, such as two successful disjuncts or two paths to the same value.
The sets describe these alternatives; a validator may return one without enumerating the others.
The returned object contains the child traces or witnesses needed to justify the check.
Table~\ref{tab:duality} gives the central cases.
\end{sloppypar}

Satisfaction traces and failure witnesses are mutually recursive through negation:
\[
\begin{split}
\mathsf{Sat}(G,v,\neg\varphi)
  &= \{\mathsf{NotHeld}(F)\mid F\in\mathsf{Fail}(G,v,\varphi)\},\\
\mathsf{Fail}(G,v,\neg\varphi)
  &= \{\mathsf{Not}(S)\mid S\in\mathsf{Sat}(G,v,\varphi)\}.
\end{split}
\]
A negated constraint passes because its child fails and fails because its child passes.
Therefore, the evidence for a negative constraint contains its child's opposite-polarity evidence.
A consumer can inspect what is required to make a child pass or fail.

\begin{table}[t]
  \caption{Evidence retained for passing and failing evaluations.
  Here $\varphi$ and $\psi$ denote two potentially different shape expressions.
  ``Support'' includes the concrete triples realizing each path.}
  \label{tab:duality}
  \vspace{-1em}
  \scriptsize
  \begin{tabularx}{\columnwidth}{@{}p{0.13\linewidth}XX@{}}
    \toprule
    Shape & $\mathsf{Sat}$ & $\mathsf{Fail}$ \\
    \midrule
    $\varphi\land\psi$
      & traces for both children
      & witnesses for failed children \\
    $\varphi\lor\psi$
      & traces for successful branches
      & witnesses for every branch \\
    $\neg\varphi$
      & $\mathsf{Fail}(\varphi)$
      & $\mathsf{Sat}(\varphi)$ \\
    $\exists_{\geq n}\pi.\varphi$
      & matched values and support
      & count deficit, qualifying and rejected values \\
    $\exists_{\leq n}\pi.\varphi$
      & checked matches and support
      & surplus values and their support \\
    \bottomrule
  \end{tabularx}
  \vspace{-2em}
\end{table}

\subsection{Canonical Evidence and Evaluation Progress}

The retention rules in Table~\ref{tab:duality} select the child evidence that establishes the parent result: a failing conjunction retains its failed children, whereas a failed disjunction retains every branch.
We call this pruned justification the \emph{canonical evidence}.
\systemname can separately attach an optional evaluation-progress record for consumers that need the status of immediate authored children, including branches omitted from canonical evidence; this record is excluded from our measurements.

\begin{figure}[t]
  \centering
  \includegraphics[width=\columnwidth]{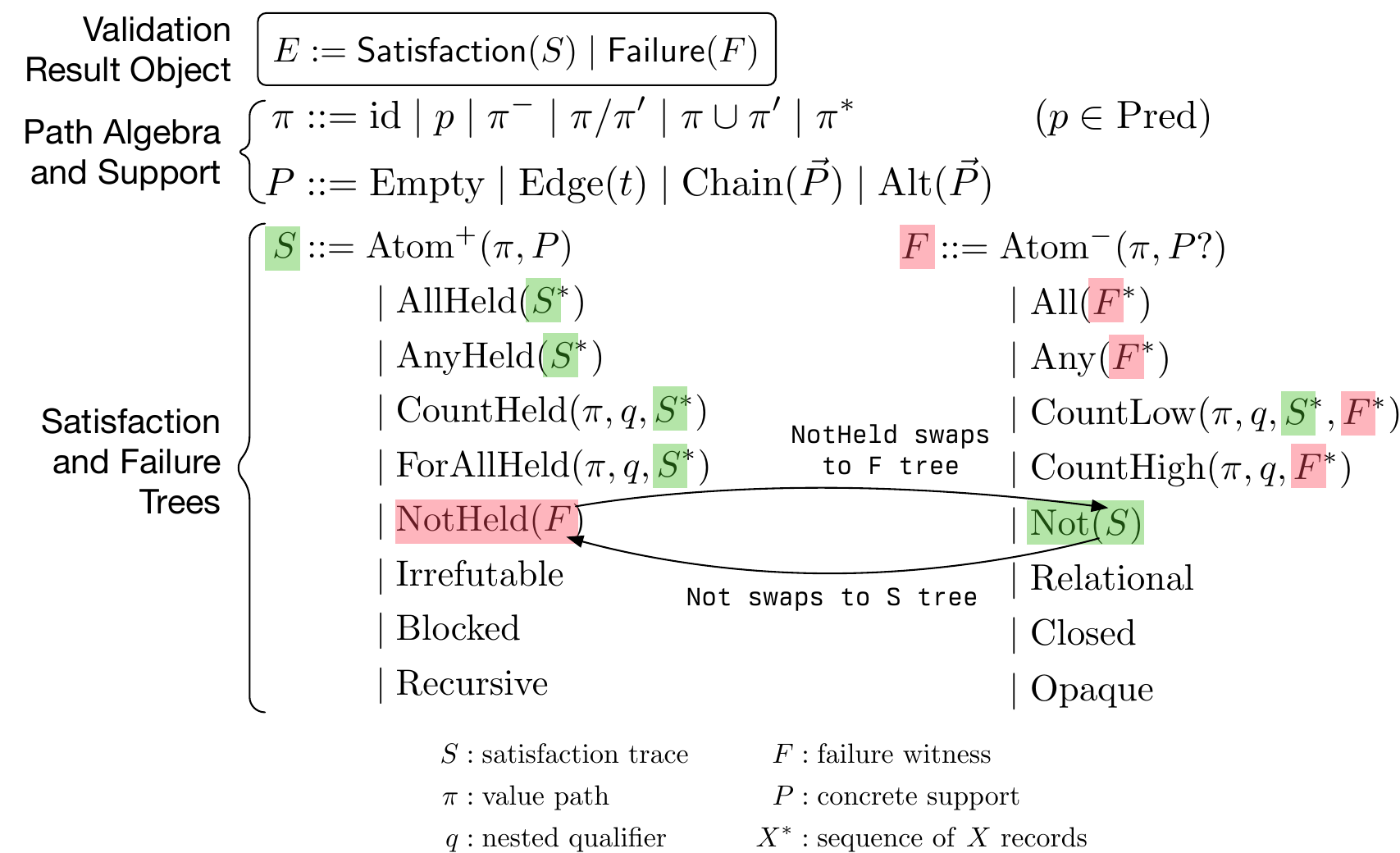}
  \vspace{-2em}
  \Description{The evidence grammar pairs a satisfaction trace and a failure witness with algebraic paths and concrete path support.
  Negation crosses between the two evidence polarities.}
  \caption{The evidence grammar.
  Negation reverses polarity on $\mathsf{NotHeld}(F)$, $\mathsf{Not}(S)$.
  A low-count witness retains traces for qualifying matches and witnesses for rejected candidates.}
  \label{fig:evidence-grammar}
  \vspace{-1em}
\end{figure}

Figure~\ref{fig:evidence-grammar} summarizes the typed structures.
Each record names its algebraic constraint; atomic records retain observed and expected values, while count records retain their path, nested condition, values, bounds, and child evidence.
A low-count witness includes both qualifying matches and rejected candidates.

\subsection{Concrete Path Support and Recursion}
Evaluating path $\pi$ over graph $G$ yields a relation $\llbracket\pi\rrbracket_G$ on RDF terms~\cite{ahmetaj_common_2025}, but the pair $(v,u)\in\llbracket\pi\rrbracket_G$ does not identify which triples connect $v$ to $u$.
Whenever evidence retains such a value $u$, it also records path support $P$:
\[
  P ::= \mathsf{Empty}
    \mid \mathsf{Edge}(t)
    \mid \mathsf{Chain}(\vec P)
    \mid \mathsf{Alt}(\vec P).
\]
$\mathsf{Empty}$ certifies the identity path, $\mathsf{Edge}(t)$ names a triple $t\in G$, $\mathsf{Chain}$ orders support for a sequence or repetition, and $\mathsf{Alt}$ groups retained routes.
When several paths connect the same endpoints, the evidence may record only one of them; deleting the recorded triples therefore need not disconnect the value, since unrecorded paths may remain.

We adopt the stratified greatest-fixpoint interpretation of recursive shape definitions formalized by Ahmetaj et al.~\cite{ahmetaj_recursive_2026}.
When validation follows a recursive shape reference back to the same node and shape, the evidence records a back-reference instead of expanding forever.
Stratification excludes cycles through negation, which our implementation rejects.

\subsection{Evidence for an HVAC Application Contract}
\begin{figure*}[t]
  \centering
  \includegraphics[width=.9\textwidth]{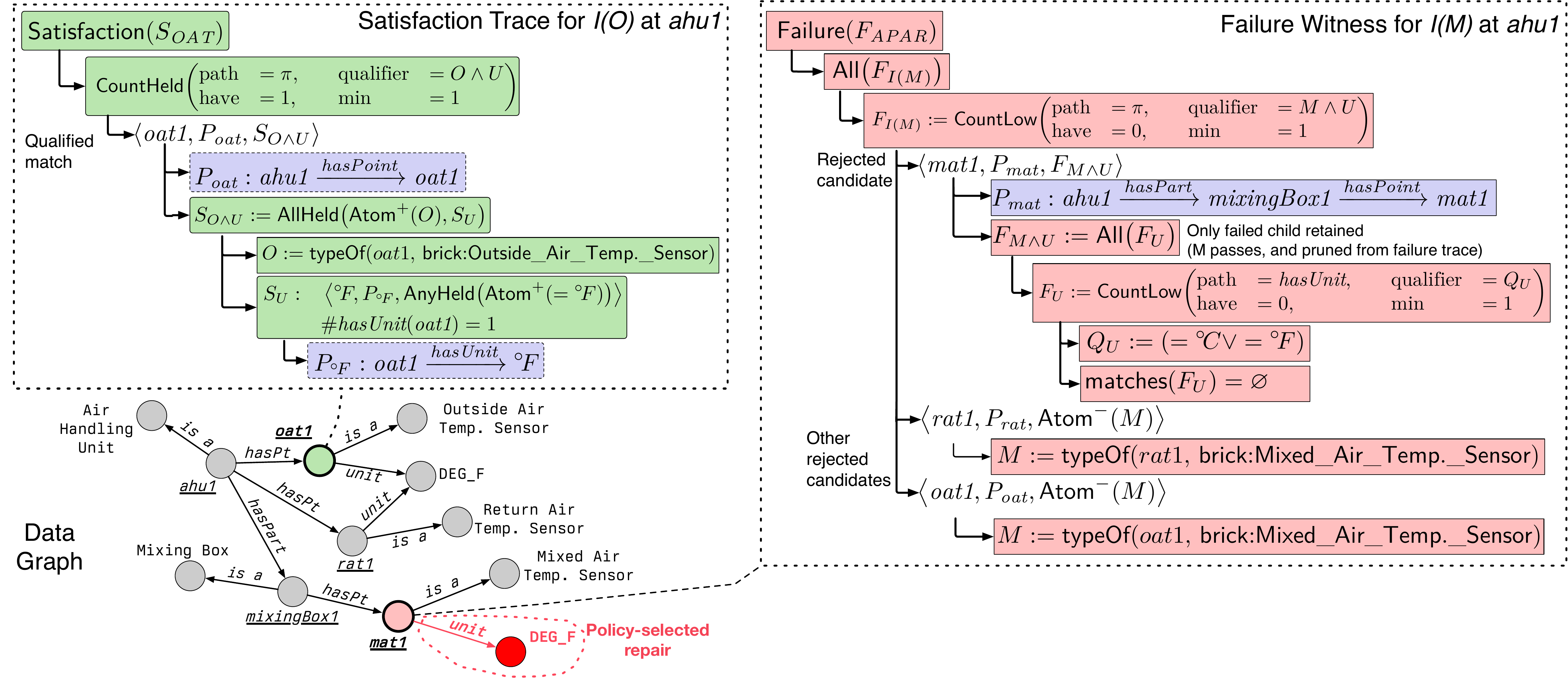}
  \vspace{-1em}
  \Description{A data graph and two connected evidence trees for ahu1.
  The satisfaction trace connects the outside-air check to oat1, its direct hasPoint edge, and its Fahrenheit unit edge.
  The failure witness connects the mixed-air check to reachable mat1, retains only the failed unit child, and records a count deficit and rejected candidates.
  A dashed red edge from mat1 to degrees Fahrenheit marks the repair selected from passing context.}
  \caption{Graph-grounded evidence for the APAR contract.
  The OAT trace connects $\mathit{ahu1}$ to a qualifying sensor and unit.
  The MAT witness retains the reachable sensor, prunes its passing type check, and localizes failure to the missing unit; passing traces support the policy-selected Fahrenheit repair.}
  \label{fig:apar-evidence}
  \vspace{-1em}
\end{figure*}

To make the evidence semantics concrete, we consider semantic building models used to configure fault-detection and optimal-control applications~\cite{pritoni_metadata_2021}.
The APAR ruleset~\cite{schein_rule-based_2006} is widely implemented in building management systems to detect common HVAC faults.
We express the inputs required by APAR Rule~1 as a shape over a Brick~\cite{balaji_brick_2018} knowledge graph.
The shape requires mixed, return, and outside air temperature sensors, which may attach directly to an air-handling unit (AHU) or to a nested component such as its mixing box.
Let $=a$ test whether the current value is the RDF term $a$, and let $M$, $R$, and $O$ test if a node is a mixed-, return-, or outside-air temperature sensor, respectively.
For a sensor-type test $T\in\{M,R,O\}$, define
\[
\begin{split}
  \pi &:= \mathit{hasPart}^{*}/\mathit{hasPoint},\\
  U &:= \exists_{\geq 1}\mathit{hasUnit}.(={}^{\circ}\!C\lor={}^{\circ}\!F)
       \land \exists_{\leq 1}\mathit{hasUnit}.\top,\\
  I(T) &:= \exists_{\geq 1}\pi.(T\land U),\\
  \mathit{APARReady} &:= I(M)\land I(R)\land I(O).
\end{split}
\]
Figure~\ref{fig:apar-evidence} places the data graph beside evidence for two checks at target node $\mathit{ahu1}$.
The graph contains complete return- and outside-air temperature points in degrees Fahrenheit, while mixed-air point $\mathit{mat1}$ lacks a unit.
Path $\pi$ reaches $\mathit{oat1}$ directly through $\mathit{hasPoint}$ and reaches $\mathit{mat1}$ through one $\mathit{hasPart}$ step followed by $\mathit{hasPoint}$.

The trace for $I(O)$ retains $\mathit{oat1}$, its direct $\mathit{hasPoint}$ edge, and its Fahrenheit $\mathit{hasUnit}$ edge.
The failure witness for $I(M)$ records a count of zero against a lower bound of one.
It also records that $\mathit{mat1}$ is reachable through the mixing box but fails only the unit requirement.
The witness thus distinguishes incomplete metadata on the intended point from missing AHU topology and from points with the wrong role.

Both possible units satisfy the missing obligation, so validation alone cannot choose between them.
We author the three input requirements as separate statements within the $\mathit{APARReady}$ shape, so their individual results preserve passing context that the canonical witness for the failed conjunction would omit.
A consumer can locate the incomplete point from the failure, extract units from the passing statements, and choose the edit supported by the building's Fahrenheit convention (Figure~\ref{fig:repair-consumer}).

\section{Realizing Evidence in \systemname}
\label{sec:system}

\systemname is an experimental SHACL and SHACL-AF engine organized around the algebra in \S\ref{sec:witnesses}, implemented in Rust with a Python interface through PyO3.

\subsection{Preparing Shapes and Data}

\systemname lowers paths, constraints, target selectors, and SHACL-AF rules into its algebra while recording authored identities.
Semantics-preserving normalization simplifies expressions, eliminates unreachable algebra nodes, canonicalizes paths, and shares identical acyclic subexpressions.

Several authored statements may normalize to the same constraint and target selection.
\systemname evaluates the resulting node--shape checks once but returns a statement record for each authored statement.
Nested evidence refers to normalized algebra nodes by identifier; a run-level \emph{constraint catalog} maps each identifier to its node kind and expression.
After SHACL-AF rules reach a least fixed point, conformance checking and evidence construction use the same normalized constraints and indexed inferred graph.


\subsection{Consumer-Facing Validation Results}

The Python interface binds an immutable shapes and data snapshot in an \texttt{Evidence}\allowbreak\texttt{Session}.
Calling \texttt{validate()} returns an \texttt{Evidence}\allowbreak\texttt{Run} grouped by authored statement.
Each statement record retains authored and normalized identifiers and contains every focus node selected by its target; an empty \texttt{selected\_foci} list records that the target selected nothing.
Each selected focus has exactly one tagged \texttt{Satisfaction} or \texttt{Failure} object.

A consumer walks the run statement by statement or looks up one focus node or named shape, then projects the fields a task needs from each result.
From a satisfaction trace, an application projects the conforming data it consumes: matched values and their supporting triples.
From a failure witness, an operator or repair program projects the missing obligations and offending values that identify the incomplete or invalid data.
The same objects render human-readable explanations, so programs read typed fields instead of parsing text.

\subsection{Efficient Evidence Materialization}
\begin{sloppypar}
The four producing calls in Table~\ref{tab:python-api} evaluate the same selected node--shape checks over the prepared snapshot; they differ in when and how much evidence they materialize.
\texttt{validate\_conformance()} retains only verdicts and counts.
\texttt{validate()} materializes eagerly, constructing traces and witnesses for both polarities during evaluation.
Eager construction follows the retention rules of Table~\ref{tab:duality}: \systemname memoizes Boolean judgments and builds only the branches canonical evidence requires, rather than retaining every evaluator call.
\texttt{find\_failures()} defers materialization, recording a handle---normalized statement and focus node---for each failing pair without constructing its witness.
\texttt{explain(pair)} then re-evaluates the one check its handle names, this time constructing canonical evidence and skipping target selection, which the handle has already resolved.
Deferral is sound because the session snapshot is immutable: re-evaluation reproduces exactly the evidence eager materialization would have built, so consumers pay for evidence only on the failures they explain.
When authored statements share a normalized node--shape pair, an on-demand explanation evaluates the pair once and returns the corresponding authored statement records.
\end{sloppypar}

\begin{table}[t]
  \caption{Python calls for producing and consuming validation evidence.}
  \vspace{-1em}
  \label{tab:python-api}
  \scriptsize
  \setlength{\tabcolsep}{3pt}
  \begin{tabularx}{\columnwidth}{@{}
      >{\raggedright\arraybackslash}p{0.39\columnwidth}
      >{\raggedright\arraybackslash}p{0.34\columnwidth}
      >{\raggedright\arraybackslash}X@{}}
    \toprule
    Call & Returned result & Consumer use \\
    \midrule
    \multicolumn{3}{@{}l}{\emph{Produce evidence over one prepared snapshot}} \\
    \texttt{validate\_conformance()}
      & Verdict and counts; no evidence
      & Check graph status \\
    \texttt{find\_failures()}
      & Counts and failing-pair handles
      & Locate failed checks \\
    \texttt{explain(pair)}
      & Pair-scoped \texttt{EvidenceRun}
      & Diagnose one check \\
    \texttt{validate()}
      & All-pair \texttt{EvidenceRun}
      & Consume both polarities \\
    \addlinespace[2pt]
    \multicolumn{3}{@{}l}{\emph{Project an \texttt{EvidenceRun}}} \\
    \texttt{failures\_for(focus)} / \texttt{satisfactions\_for(focus)}
      & Polarity-specific evidence lists
      & Select relevant checks \\
    \texttt{offending\_values()} / \texttt{missing\_obligations()}
      & Values and cardinality deficits
      & Diagnose failed checks \\
    \texttt{values\_for\_path(path)}
      & Matched values on one path
      & Extract conforming data \\
    \bottomrule
  \end{tabularx}
  \vspace{-2em}
\end{table}

\section{Evaluation}
\label{sec:evaluation}

We benchmark 45 Brick building models (knowledge graphs) from Mortar~\cite{fierro_mortar_2018} against the shapes and rules embedded in Brick 1.5.0~\cite{balaji_brick_2018} and its transitive import closure.
We benchmark 19 ASHRAE~223P models from Open223~\cite{open223_models_2026,roth_supervisory_2022} against the June 24, 2026 ontology snapshot and its import closure.
Neither corpus uses a separate benchmark shape suite.
Because the models predate these ontology snapshots, violations include Brick deprecations and 223P corrections or changes; the failure fraction characterizes the workload, not corpus quality.
All timed modes use the same prepared shapes and indexed graph, and the harness checks that conformance-only and evidence-producing results agree.
Finally, our APAR case study exercises the validation result object as an application and repair interface.

\subsection{Evidence Materialization Cost}

\begin{figure}[t]
  \centering
  \includegraphics[width=.49\linewidth]{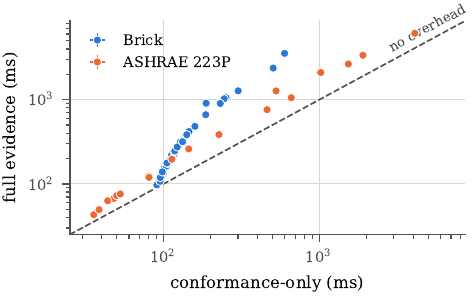}\hfill
  \includegraphics[width=.49\linewidth]{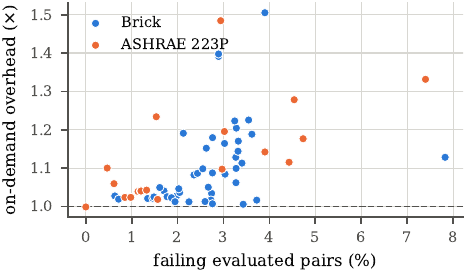}
    \vspace{-1em}
  \caption{Per-model medians; conformance-only latency is the baseline in both panels.
  Left: all-pair canonical evidence latency; error bars mark median absolute deviation over ten rounds; the dashed diagonal marks equal cost.
  Right: on-demand overhead (discovery plus explanation of every failing pair) against the failed-pair fraction; the dashed line marks equal cost.}
  \Description{Two scatter plots compare all-pair evidence latency with conformance latency and on-demand overhead with the percentage of failed pairs for Brick and ASHRAE 223P models.}
  \label{fig:evidence-results}
\end{figure}

We ran each mode ten times per model on one core of an Apple M2 MacBook Air with 24~GB of RAM.
Parsing, SHACL-AF inference, normalization, and indexing run once per model, untimed, so the ratios isolate evidence construction.
The harness rotates three modes---conformance-only validation, eager all-pair canonical evidence, and failure discovery with on-demand explanation of every failing pair---and checks that each on-demand explanation exactly matches its all-pair counterpart.
Figure~\ref{fig:evidence-results} compares each evidence mode with the conformance-only baseline.

Each model contributes the median of its paired per-round ratios; across models, all-pair materialization costs a median $2.07\times$ conformance-only execution on Brick and $1.54\times$ on 223P.
Only 3.1\% and 3.9\% of the 247,333 Brick and 228,760 223P selected node--shape checks fail, so discovering and explaining only failures reduces the ratios to $1.06\times$ and $1.10\times$.
For the 223P model with no failures, discovery costs $1.01\times$ conformance, while overhead rises with the failed-pair fraction (Figure~\ref{fig:evidence-results}, right).

\subsection{Consuming Evidence for an HVAC Contract}

The APAR contract requires mixed-, return-, and outside-air temperature inputs with declared units.
Figure~\ref{fig:repair-consumer} applies the Python projections from Table~\ref{tab:python-api} to this contract.
The consumer locates the incomplete point and cardinality deficit from the failure, then obtains the building's unit convention from the passing statements.
The consumer obtains one failure and two satisfactions for $\mathit{ahu1}$ without walking internal evaluator state.
The failure's typed projections identify $\mathit{mat1}$ as a reachable rejected candidate and report one missing value; the satisfactions project the two Fahrenheit values and their supporting paths.
The consumer proposes $(\mathit{mat1},\mathit{hasUnit},{}^{\circ}\!F)$, using the failure to locate the edit and the passing statements to choose its value.

\begin{figure}[t]
\begin{lstlisting}[style=shiftypython]
run = shifty.EvidenceSession(apar_shape, data_graph).validate()
mat, = run.failures_for(AHU1)
assert MAT1 in mat.offending_values()
assert mat.missing_obligations()[0].missing == 1
units = [unit for sat in run.satisfactions_for(AHU1)
         for unit in sat.values_for_path(HAS_UNIT)]
assert len(set(units)) == 1
unit = units[0]  # DEG_F in this graph
repair = (MAT1, HAS_UNIT, unit)
\end{lstlisting}
\vspace{-1em}
\caption{A consumer combines the failure witness with passing traces to select a repair consistent with the building's existing unit convention. Uppercase names denote RDF terms.}
\Description{Python code validates the APAR shape, locates the mixed-air point with missing metadata, extracts the unit from passing checks, and constructs a repair that assigns the same unit to the incomplete point.}
\label{fig:repair-consumer}
\vspace{-1em}
\end{figure}

\section{Future Work}
\label{sec:discussion}

Evidence-carrying validation makes result lineage part of the validator's contract: execution may reorder and share work, but it must retain enough information to construct the specified evidence.
We intend to develop this contract into a general interface between knowledge graphs and their programmatic consumers.

Semantic models of buildings and other cyber-physical systems exist to configure digitally delivered applications such as controls, analytics, and fault detection~\cite{roth_supervisory_2022,pritoni_metadata_2021}.
Platforms like BuildingMOTIF already run validation in a loop during model authoring: shapes describe application requirements, and each failure indicates what to add next~\cite{fierro_application_2022}.
The platform currently reconstructs that guidance from a failure-oriented report, recomputing facts the validator discarded.
With evidence-carrying validation, the platform can instead project the validating subgraph to configure applications and the missing obligations to guide authoring.

LLM agents that construct, query, and repair knowledge graphs are a second consumer.
Our prior evaluation of LLM-based graph repair found that repair quality depends on prompt construction: the most effective prompts combine the violated constraint with the relevant contextual fragment of the graph~\cite{lin_systematic_2025}.
Assembling that fragment is currently an ad hoc retrieval problem.
A failure witness contains the evaluated constraint, the rejected candidates, and their supporting triples, and satisfaction traces record the conventions a proposed edit should follow (\S\ref{sec:evaluation}).
We plan to evaluate evidence objects as agent context against retrieval-based baselines, and to add an application-binding layer that returns validated values under stable names instead of requiring consumers to traverse evidence.

The shape algebra~\cite{ahmetaj_common_2025} also gives the engine an optimizable intermediate representation.
\systemname lowers SHACL into the algebra before reading data, so standard database techniques apply: normalization and shared-subexpression evaluation already collapse redundant constraints across thousands of shapes, and cost-based planning, index selection, and incremental revalidation of affected focus nodes are direct analogues of query optimization.
Lowering the absolute validation cost reduces the practical impact of the overheads in \S\ref{sec:evaluation}.
The $1.5$--$2.1\times$ relative cost matters less once validation completes in (milli)seconds, meaning evidence can remain enabled inside every iteration of an authoring or agent loop.
Because application contracts and knowledge graphs evolve independently, incremental maintenance could also update affected traces and witnesses without rebuilding the complete validation run.
Complete repair semantics, external candidate providers, joint planning, and cost models remain separate research problems.

\section{Related Work}
\label{sec:related}

Outside RDF, data-quality frameworks such as Deequ validate tabular data against declarative constraints, reporting per-constraint metrics and failing records rather than per-record justification of passes and failures~\cite{schelter_automating_2018}.

Delva et al. define a SHACL neighborhood as an RDF subgraph sufficient to preserve conformance and obtain a nonconformance neighborhood by negating the shape~\cite{delva_provenance_2023}.
First-order provenance games likewise follow both sides of logical evaluation through negation~\cite{kohler_provenance_2013}.
\systemname instead returns typed validation objects that retain constraint structure, cardinality evidence, concrete path support, and recursive back-references without claiming that the projected support subgraph preserves conformance.

Rule-based RDF validation can return proof-like violation causes, and xpSHACL combines a violation tree with retrieved context to generate prose~\cite{demeester_rule_2021,publio_xpshacl_2025}.
\systemname exposes one structured interface whose passing and failing evidence these and other explainers can build on.

Database repairs formalize minimal consistent instances, and SHACL repair systems enumerate subset- or cardinality-minimal additions and deletions~\cite{arenas_consistent_1999,ahmetaj_explanations_2021,ahmetaj_repairing_2022}.
Minimality identifies smaller changes but does not determine which edit reflects the domain.
\systemname returns the validation evidence before a consumer chooses an edit, so application policy can combine failure information with passing context.
The evidence does not enumerate repairs; repair selection is one consumer of it.

\section{Conclusion}
\label{sec:conclusion}

We introduced \emph{evidence-carrying validation}: every selected node--shape check returns a satisfaction trace or a failure witness, mutually recursive objects that ground each verdict in constraints, cardinality decisions, and supporting triples.
\systemname realizes this interface for SHACL, preserving authored constraint identities through normalization and exposing typed Python projections over both polarities.
Materializing this evidence adds modest overhead to conformance-only validation and our case study shows passing and failing evidence together locating missing metadata and selecting a domain-consistent repair.
Validation that preserves what the validator learned lets applications, platforms, and agents keep evolving graphs and their constraints in step.

\bibliographystyle{ACM-Reference-Format}
\bibliography{references}

\end{document}